\documentclass{iau_FM}
\usepackage{graphicx}

\title[FM4: Chemical abundances in spirals] 
{Chemical abundances from planetary nebulae in local spiral galaxies}

\author[Michael Richer \& Marshall McCall]   
{Michael G. Richer$^1$ \and Marshall L. McCall$^2$}

\affiliation{$^1$Instituto de Astronom\'\i a, Universidad Nacional Aut\'onoma de M\'exico, Apartado Postal 106, 22800 Ensenada, Baja California, M\'exico \\ email: {\tt richer@astrosen.unam.mx}\\

$^2$Department of Physics and Astronomy, York University, Toronto, Ontario L3T 3R1, Canada  \\email: {\tt mccall@yorku.ca}}

\pubyear{2015}
\jname{Astronomy in Focus, Volume 1} 
\editors{Piero Benvenuti, ed.}
\begin{document}

\maketitle

\begin{abstract}
While the chemical abudances observed in bright planetary nebulae in local spiral galaxies are less varied than their counterparts in dwarfs, they provide new insight.  Their helium abundances are typically enriched by less than 50\% compared to the primordial abundance.  Nitrogen abundances always show some level of secondary enrichment, but the absolute enrichment is not extreme.  In particular, type I PNe are rare among the bright PNe in local spirals.  The oxygen and neon abundances are very well correlated and follow the relation between these abundances observed in star-forming galaxies, implying that either the progenitor stars of these PNe modify neither abundance substantially or that they modify both to maintain the ratio (not predicted by theory).  According to theory, these results imply that the progenitor stars of bright PNe in local spirals have masses of about $2\,\mathrm M_{\odot}$ or less.  If so, the progenitors of these PNe have substantial lifetimes that allow us to use them to study the recent history of their host galaxies, including gravitational interactions with their neighbours.  Areas that require further study include the systematic differences observed between spectroscopy obtained through slits and fibres, the uncertainties assigned to chemical abundances, including effects due to ionization correction factors, and the physics that gives rise to the PN luminosity function.  Indeed, so long as we lack an understanding of how the last arises, our ability to use bright PNe as probes to understand the evolution of their host galaxies will remain limited.  
\keywords{planetary nebulae: general, stars: evolution, ISM: abundances, Galaxy: evolution}
\end{abstract}

\firstsection 
\section{Introduction}

In the past few years, substantial studies of chemical abundances in planetary nebulae (PNe) in the discs of local spirals have begun to appear.  So far, the galaxies observed include only M31, M33, NGC 300, and M81.  These observations lagged behind their counterparts in the bulge of M31, Cen A, and dwarf galaxies due to the challenge of separating them from the numerous compact H {\sc ii} regions and the diffuse ionized gas that pervades spiral discs (see Fig. \ref{fig1} for a list of the data).  Given the area of spiral discs, multi-object spectrographs, employing both slits and fibres, have had to be used to obtain the required data.  The reader should keep in mind that the PNe observed thus far are all among the brightest PNe in their galaxies, and it is not yet clear that these necessarily sample the entire population of PNe or the entire population of stars that give rise to PNe.  For a complementary view of the bright PNe in dwarf galaxies, see Gon\c calves (this volume) or \cite[Gon\c calves et al. (2014)]{goncalvesetal2014}.

\section{Results}

\begin{figure}[]
\begin{center}
 \includegraphics[width=0.5\columnwidth]{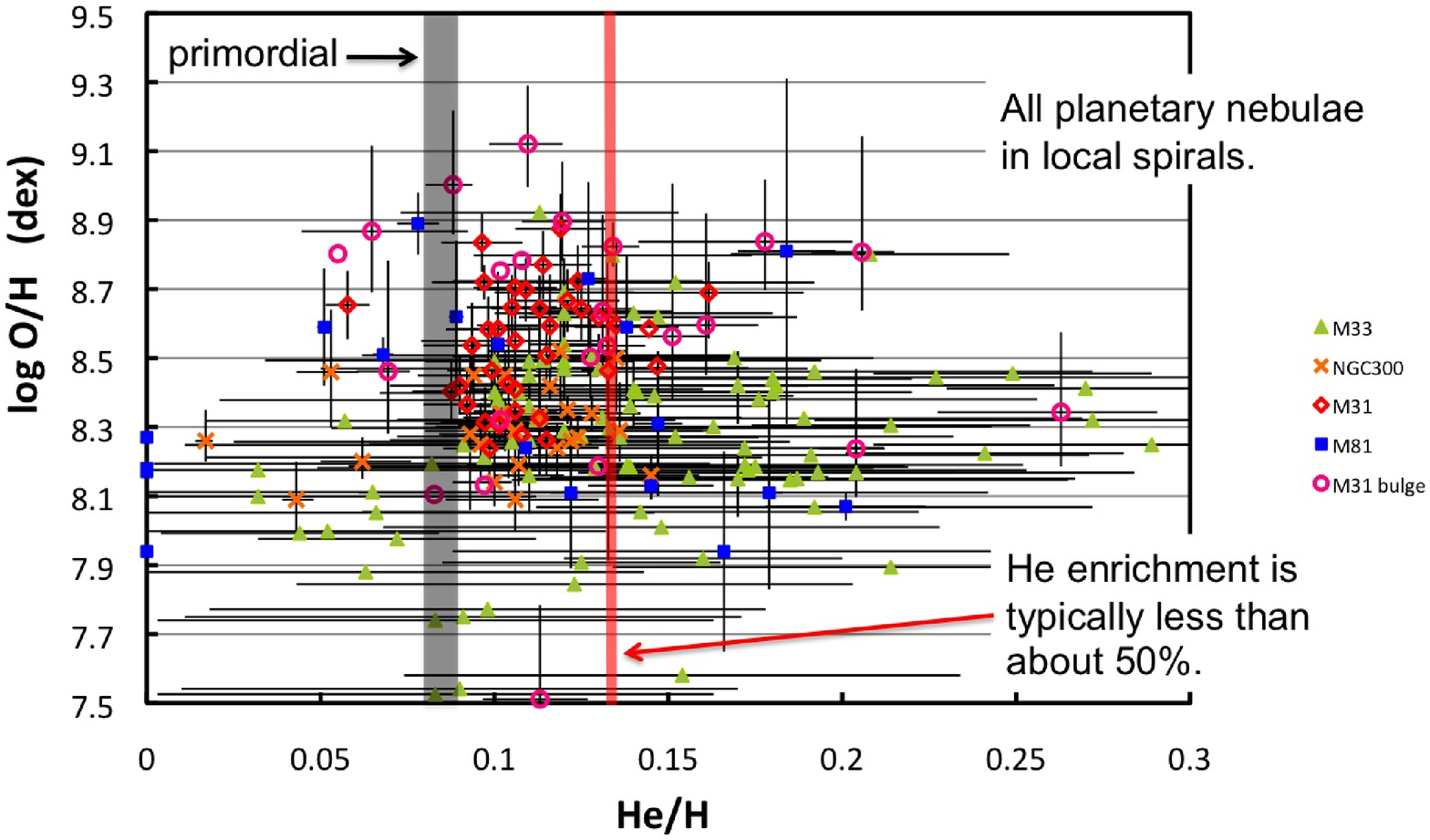} \\
 \includegraphics[width=0.49\columnwidth]{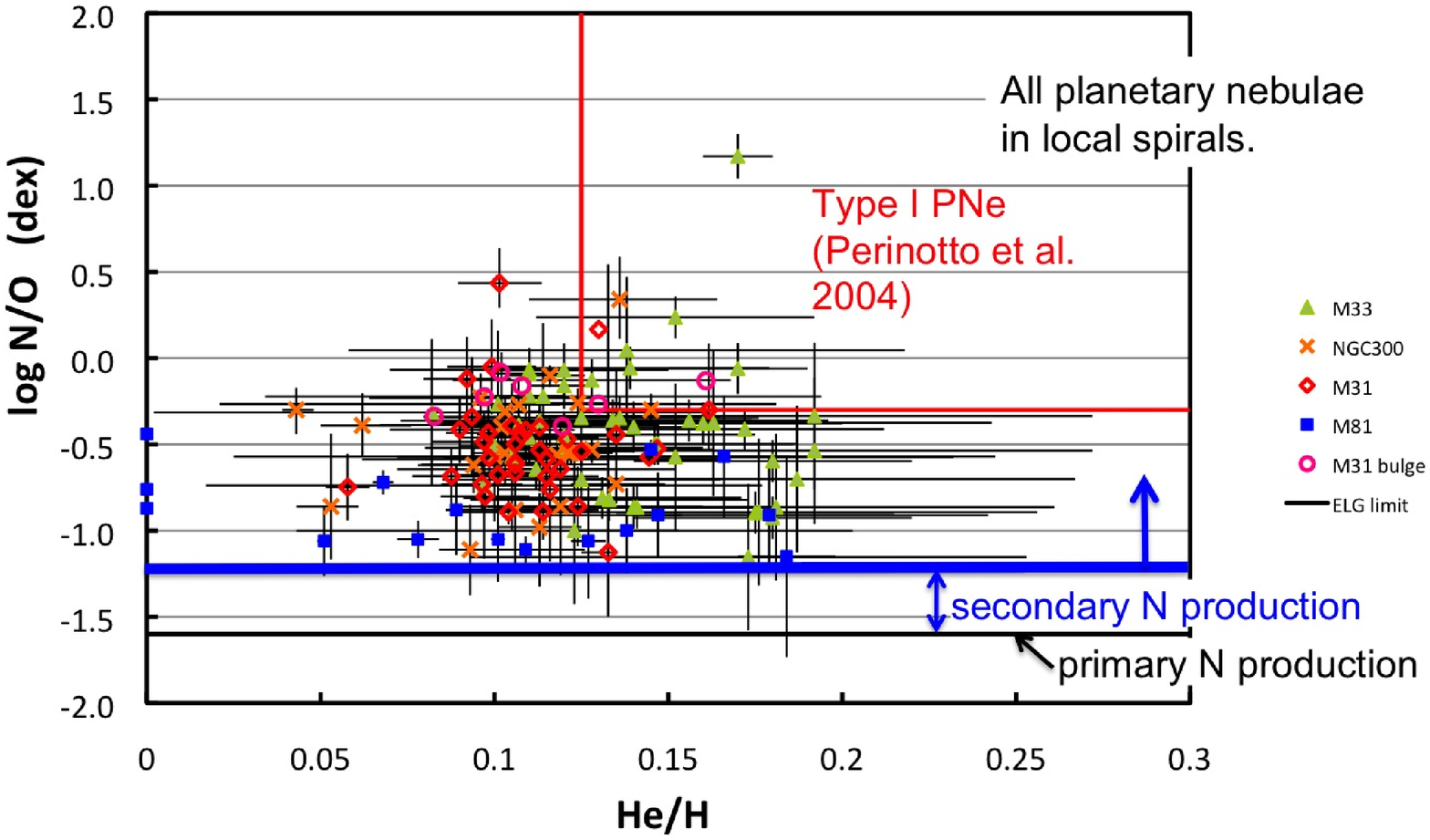} 
 \includegraphics[width=0.49\columnwidth]{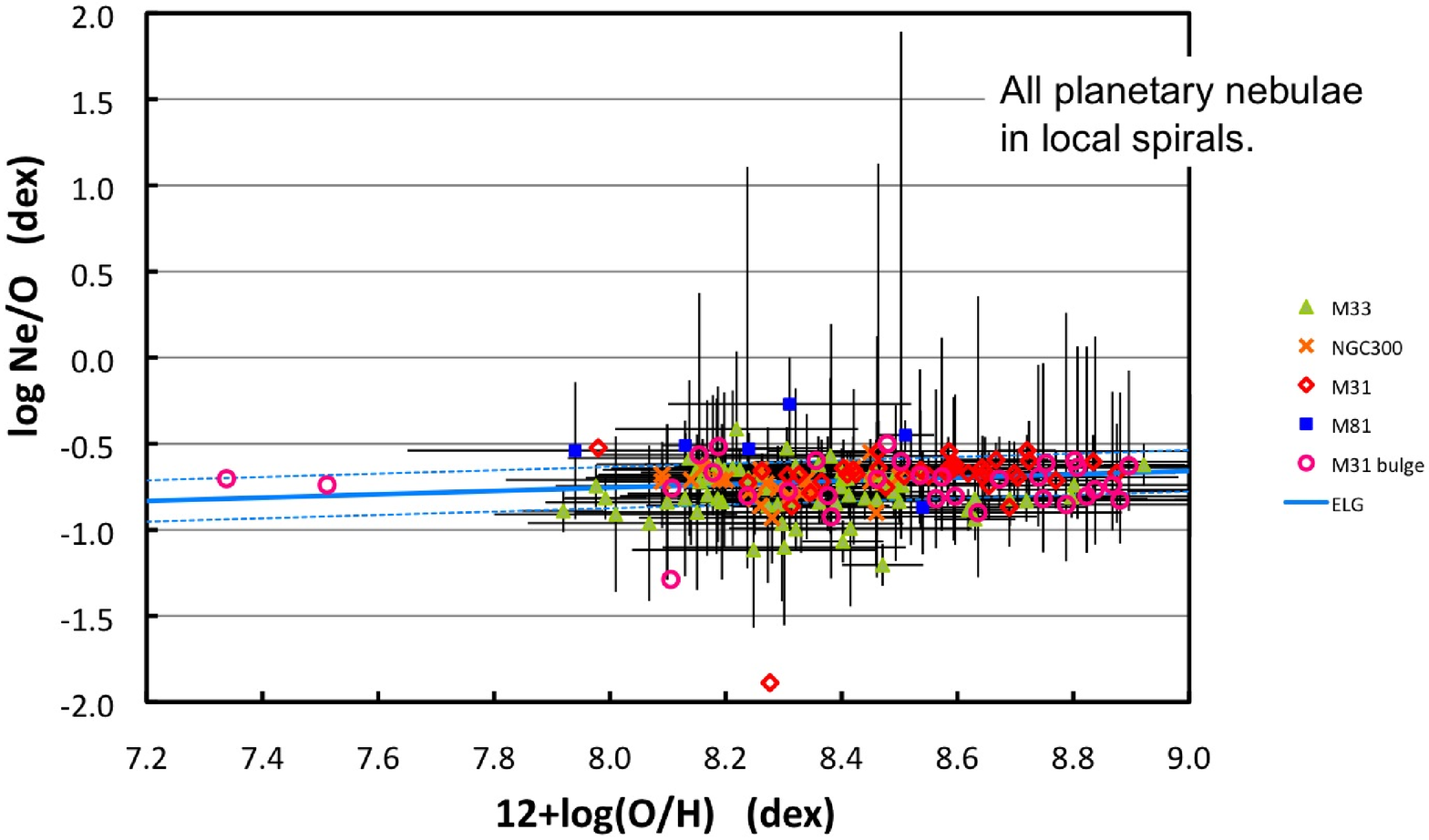} 
 \caption{We summarize the results regarding chemical abundances in bright PNe in local spiral galaxies, including the bulge of M31.  The galaxies are identified in the legends.  In general, bright PNe in local spirals show modest enrichment in He, essentially all show some level of secondary nitrogen enrichment, though Type I PNe are rare, and the Ne and O abundances are very well correlated.  The definition of Type I PNe (red lines, bottom left panel) is that of \cite[Perinotto et al. (2004)]{perinottoetal2004} for the Milky Way.  As regards Ne/O, the great majority of the data points and virtually all of the error bars lie within the uncertainty about the mean relation between Ne/O and O/H for star-forming galaxies from \cite[Izotov et al. (2006)]{izotovetal2006}.  The definition of primary enrichment of N also comes from this reference.  Data sources for all analyses:  {\bf M31} \cite[Jacoby \& Ford (1986)]{jacobyford1986}; \cite[Richer et al. (1999)]{richeretal1999}; \cite[Jacoby \& Ciardullo (1999)]{jacobyciardullo1999}; \cite[Roth et al. (2004)]{rothetal2004}; \cite[Kwitter et al. (2012)]{kwitteretal2012}; \cite[Sanders et al. (2012)]{sandersetal2012}; \cite[Fang et al. (2013)]{fangetal2010}; \cite[Balick et al. (2013)]{balicketal2013}; \cite[Kniazev et al. (2014)]{kniazevetal2014}; \cite[Corradi et al. (2015)]{corradietal2015} {\bf M33} \cite[Magrini et al. (2009)]{magrinietal2009}; \cite[Bresolin et al. (2010)]{bresolinetal2010} {\bf NGC 300} \cite[Stasi\'nska et al. (2013)]{stasinskaetal2013} {\bf M81} \cite[Stanghellini et al. (2010)]{stanghellinietal2010}; \cite[Stanghellini et al. (2014)]{stanghellinietal2014}; \cite[Richer \& McCall (2015)]{richermccall2015}.}
   \label{fig1}
\end{center}
\end{figure}

Fig. \ref{fig1} summarizes the results obtained thus far for the abundances of the elements helium, nitrogen, oxygen, and neon.  (The scarcity of data for argon and sulphur does not warrant a detailed analysis.)  The top panel presents $12+\log(\mathrm O/\mathrm H)$ versus $\mathrm {He}/\mathrm H$.  Most of the bright PNe in local spirals have neither very low nor very high oxygen abundances, with most falling within the range $7.9\,\mathrm{dex} < 12+\log(\mathrm O/\mathrm H) < 8.9\,\mathrm{dex}$.  This is not unexpected given past observations or theoretical expectations (\cite[Dopita et al. 1992, Richer \& McCall 2008]{dopitaetal1992, richermccall2008}).  This may mostly reflect that the chemical evolution of the discs of spirals is slow.  The bulge of M31 shows a greater range of oxygen abundances than the spiral discs.  

In the top panel of Fig. \ref{fig1}, the primordial helium abundance (e.g., \cite[Peimbert et al. 2007, Izotov et al. 2014]{peimbertetal2007, izotovetal2014}) as well as a value 50\% larger are shown, between which most of the measurements lie.  Since most of the progenitor stars of these bright PNe were formed with a helium abundance above the primordial value, the enrichment of the helium abundance in the envelope during the evolution of these stars must have been less than 50\%.  

The left panel in Fig. \ref{fig1} presents the $\mathrm N/\mathrm O$ abundance ratio as a function of $\mathrm {He}/\mathrm H$.  The top right part of this diagram defines the region in which type I PNe within the Milky Way are found (\cite[Perinotto et al. 2004]{perinottoetal2004}).  Clearly, the progenitor stars of all bright PNe in local spirals formed from gas that had suffered some level of secondary enrichment, because the lower envelope to the data exceeds the expectation for primary production of nitrogen by type II supernovae as judged from observations in metal-poor star-forming galaxies (\cite[Izotov et al. 2006]{izotovetal2006}).  With respect to the secondary enrichment level, the stellar progenitors of these PNe further enriched their envelopes during their evolution, typically by less than a factor of 10 (1.0\,dex), but even so a challenge for models.  

Finally, the right panel in Fig. \ref{fig1} illustrates the excellent correlation between oxygen and neon abundances.  The solid line is the relation between oxygen and neon abundances observed in star-forming galaxies (\cite[Izotov et al. 2006]{izotovetal2006}), the result of the yields from type II supernovae.  Evidently, if the progenitors of bright PNe in local spirals follow the same relation, either they modify neither abundance or modify both so as to maintain the proportion seen in the interstellar medium of star-forming galaxies.  

\section{Theoretical Interpretation}

We compare these observations with the models of \cite[Karakas (2010; henceforth K10)]{karakas2010}, \cite[Cristallo et al. (2011; C11)]{cristalloetal2011}, and \cite[Pignatari et al. (2013; P13)]{pignatarietal2013}.  While these models may not span the entire parameter space available, they are indicative of the results from models with conservative implementations of semi-convection.  At the risk of over-generalizing, their helium production is similar and modest.  K10 produces somewhat more nitrogen than the other models, especially for masses below 3-4\,M$_{\odot}$, but less neon and much less oxygen (some K10 models even consume oxygen).  P13 produce substantially more oxygen than K10 or C11.  All of these models are able to produce type I PNe only for $\mathrm M> 5-6$\,M$_{\odot}$ and K10 can exceed the enrichment observed.  We recall that the relevant metallicity range for the observations presented here is $Z\approx 0.004-0.02$.

The observation that is most decisive concerning the mass of the progenitor stars is the tight correlation of oxygen and neon abundances.  The mass above which the $(\mathrm{Ne}/\mathrm O)$ ratio departs from the observed value is approximately $2.1-2.25$\,M$_{\odot}$ ($Z=0.004-0.02$) and 2.0\,M$_{\odot}$ ($Z=0.003-0.02$), for K10 and C11, respectively.  The P13 models at $Z=0.01-0.02$ are marginally compatible with the observed relation, but none of their models are compatible at lower metallicities (due to oxygen production).   The rarity of type I PNe imply that high mass stellar progenitors produce very few bright PNe.  Therefore, these model results indicate that the stellar progenitors of bright PNe in local spirals typically have masses of about 2.0\,M$_{\odot}$ or less.  

Interpreting nitrogen abundances is complicated (e.g., \cite[Richer \& McCall 2008]{richermccall2008}).  No models can increase $\mathrm N/\mathrm O$ by an order of magnitude and maintain the tight correlation of the oxygen and neon abundances.  

\section{Concerns}

When comparing the oxygen abundances from PNe and H {\sc ii} regions in a given galaxy, it is common to find a greater dispersion among the oxygen abundances from the PNe.  If this dispersion were only to lower values for the PNe, it would be less worrisome, since PNe arise from progenitors that were formed in the past, when metallicities were lower.  However, it is also common to find many high values, above the mean observed for the local H {\sc ii} regions.  The oxygen abundances from fibre spectroscopy in M33 are the most extreme example, with PNe having oxygen abundances in excess of the largest observed in its H {\sc ii} regions, even when uncertainties are considered (\cite[Stasi\'nska et al. 2013]{stasinskaetal2013}).

A possible explanation is that the progenitors of bright PNe synthesize oxygen.  If so, they 
must also synthesize neon proportionately (\S 2).  As \cite[Karakas \& Lattanzio (2003)]{karakaslattanzio2003} point out, there is no reason that the progenitors of PNe should synthesize oxygen and neon in the ratio observed in the interstellar medium in star-forming galaxies, which is set by the nucleosynthesis of type II supernovae.  While the progenitors of PNe can synthesize oxygen and neon, the nucleosynthetic processes are different and, in the case of neon, a different isotope is produced.  It is also possible that the analysis is incomplete and that the real uncertainties are underestimated.  In the case of fibre spectroscopy of M33, emission from diffuse ionized gas  may contaminate [O{\sc ii}] $\lambda$3727 from its PNe.
\begin{figure}[]
\begin{center}
 \includegraphics[width=0.49\columnwidth]{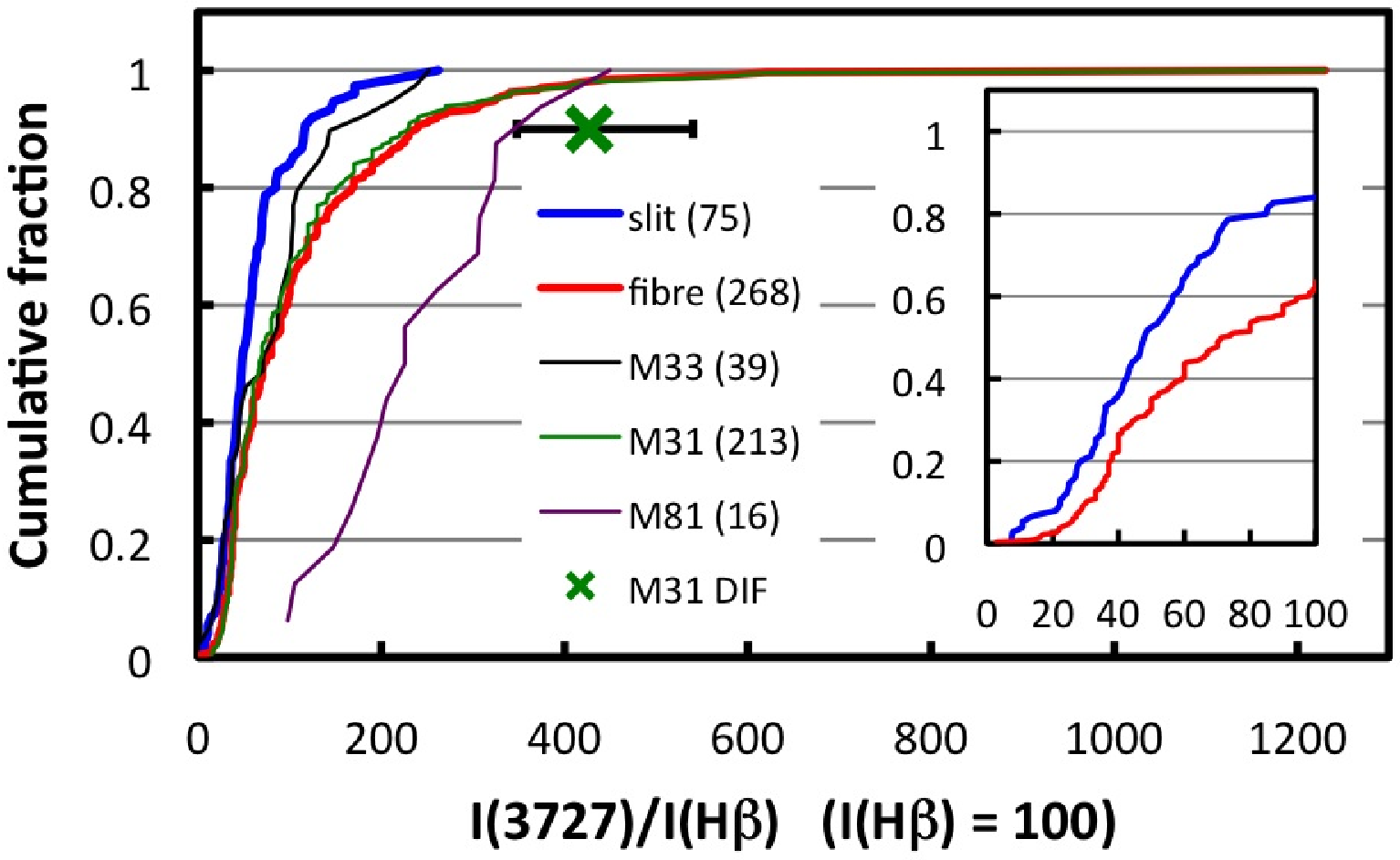} 
 \includegraphics[width=0.49\columnwidth]{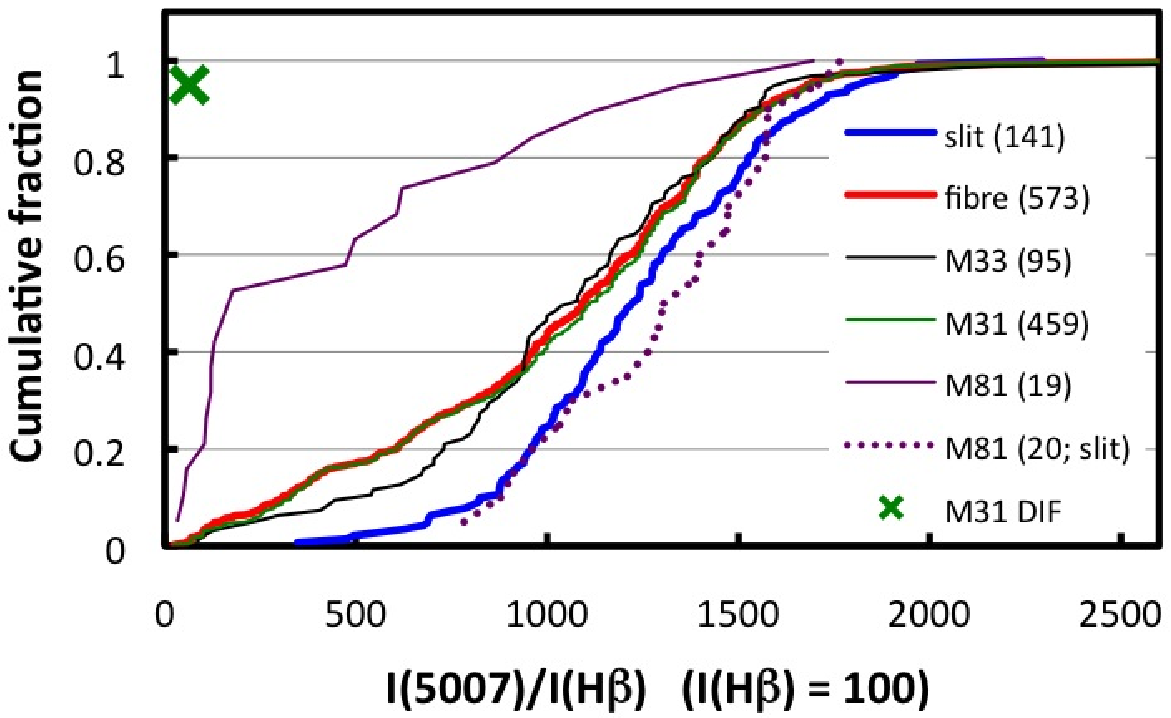} 
\end{center}
 \caption{We present the cumulative distribution functions of the intensities of [O {\sc ii}] $\lambda$3727 and [O {\sc iii}] $\lambda$5007 relative to H$\beta$ (intensity of H$\beta$ normalized to 100).  The ``fibre" and ``slit" distributions amalgamate all measurements of each type for all galaxies.  For each galaxy with fibre spectroscopy, the distributions are shown separately as well.  The green cross represents a typical line ratio (or range) for the diffuse ionized gas in the disc of M31 (\cite[Greenawalt et al. 1997]{greenawaltetal1997}) and is placed arbitrarily in the vertical direction.  The inset in the left panel shows that the difference in the distribution of the [O {\sc ii}] $\lambda$3727 intensities persists to the lowest values.  The distribution of [O {\sc ii}] $\lambda$3727 and [O {\sc iii}] $\lambda$5007 line intensities between slit and fibre spectroscopy is in the sense that might be explained by contamination from diffuse ionized gas in the fibre measurements (non-local sky subtraction), being worse for M81 than for M31 or M33, perhaps because it is the most distant system.  Note that the slit spectroscopy for PNe in M81's disc (right panel) does not differ from the slit spectroscopy for other systems.  The difference between the slit and fibre distributions of the [O {\sc ii}] $\lambda$3727 and [O {\sc iii}] $\lambda$5007 line intensities is significant, with probabilities of only $2.0\times 10^{-5}$ and $1.9\times 10^{-6}$, respectively, of arising from the same parent distribution.}
   \label{fig2}
\end{figure}

Fig. \ref{fig2} presents the observed distributions of [O {\sc ii}] $\lambda$3727 and [O {\sc iii}] $\lambda$5007 line intensities relative to H$\beta$ for bright PNe observed through slits and fibres.  There is a clear statistical difference between the distributions for both lines.  In general, slit spectroscopy finds weaker [O {\sc ii}] $\lambda$3727 and stronger [O {\sc iii}] $\lambda$5007 line intensities than fibre spectroscopy.  This might be due to the difficulty of subtracting the diffuse ionized gas that pervades the discs of spirals from the fibre spectroscopy, since this ionized gas has strong [O {\sc ii}] $\lambda$3727 and weak [O {\sc iii}] $\lambda$5007 as compared with a typical PN spectrum.  This needs further study so that the difference may be understood.

A third concern is the uncertainty associated with chemical abundances.  In the top panel in Fig. \ref{fig1}, most of the data points that nominally fall below the primordial helium abundance are compatible with this abundance, but some are not.  Presumably, this implies problems with the scheme to compute the total elemental abundance, likely the ionization correction factor employed to convert the observed ionic abundances into a total elemental abundance.  It would be helpful to investigate what uncertainty the correction for unseen ionization stages introduces, for all elements.  \cite[Delgado-Inglada et al. (2015)]{delgadoingladaetal2015} provide a very helpful introduction to this problem.  

Finally, the PNe in the disc of M31 make a very compelling case for paying close attention to the history of the galactic host.  As \cite[Kwitter et al. (2012)]{kwitteretal2012}, \cite[Balick et al. (2013)]{balicketal2013}, and \cite[Corradi et al. (2015)]{corradietal2015} argue, it appears that a substantial fraction of the PNe observed in the outer disc of M31 were not formed there, but that their progenitors were scattered to their current locations as a result of gravitational interactions between M31 and M33.  In M31, data exist for the resolved stellar content that leaves open this possibility.  In most other galaxies, such data do not exist.  If the progenitor stars of bright PNe have masses of 2\,M$_{\odot}$, they live long enough that they may feel the evolution of their host galaxy, so it is important to keep this possibility in mind when interpreting results, e.g., gradients in chemical abundances.  

\section{Conclusions}

The chemical abundances in bright PNe in local spirals turn out to be somewhat less varied than their counterparts in dwarf galaxies.  Helium abundances are enriched by a modest amount, usually less than 50\% with respect to the primordial value.  Nitrogen abundances span a wider range of enrichments than models are able to accomodate while preserving other abundance ratios, but type I PNe are rare.  Oxygen and neon abundances are very strongly correlated and closely follow the relation observed between these elements in the interstellar medium in star-forming galaxies (\cite[Izotov et al. 2006]{izotovetal2006}).  Using models to interpret these results, we deduce that the progenitors of these bright PNe have masses of about 2\,M$_{\odot}$ or less, with the relation between oxygen and neon abundances providing the strongest constraint.  If so, the progenitor stars of bright PNe in local spirals live long enough that they \emph{will} reflect the history of their host galaxies, including gravitational interactions with their neighbours.  Indeed, ensembles of bright PNe in local spirals should be good probes of both the evolution of their progenitor stars and the recent history (1-2\,Gyr) of their galactic hosts.


\begin{thebibliography}{}

\bibitem [Balick et al. (2013)]{balicketal2013} Balick, B., Kwitter, K. B., Corradi, R. L. M., et al. 2013, \textit{ApJ}, 774, 3

\bibitem [Bresolin et al. (2010)]{bresolinetal2010} Bressolin, F., Stasi\'nska, G., V\'\i lchez, J. M., et al. 2010, \textit{MNRAS}, 404, 1679

\bibitem [Corradi et al. (2015)]{corradietal2015} Corradi, R. L. M., Kwitter, K. B., Balick, B., et al. 2015, \textit{ApJ}, 807, 181

\bibitem [Cristallo et al. (2011)]{cristalloetal2011} Cristallo, S., Piersanti, L., Straniero, O. et al. 2011, \textit{ApJS}, 197, 17

\bibitem [Delgado-Inglada et al. (2015)]{delgadoingladaetal2015} Delgado-Inglada, G., Morisset, C., \& Stasi\'nska, G. 2015, \textit{MNRAS}, 440, 536

\bibitem [Dopita et al. (1992)]{dopitaetal1992} Dopita, M. A., Jacoby, G. H., \& Vassiliadis, E. 1992, \textit{ApJ}, 389, 27

\bibitem [Fang et al. (2013)]{fangetal2010} Fang, X., Zhang, Y., Garc\'\i a-Benito, R., et al. 2013, \textit{ApJ}, 774, 138

\bibitem [Gon\c calves et al. (2014)]{goncalvesetal2014} Gon\c calves, D. R., Magrini, L., \& Teodorescu, A. M. 2014, \textit{MNRAS}, 444, 1705

\bibitem [Greenawalt et al. 1997]{greenawaltetal1997} Greenawalt, B., Walterbos, R. A. M., \& Braun, R. 1997, \textit{ApJ}, 483, 666

\bibitem [Izotov et al. (2006)]{izotovetal2006} Izotov, Y. I., Stasi\'nska, G., Meynet, G., et al. 2006, \textit{A\&A}, 448, 955

\bibitem[Izotov et al. (2014)]{izotovetal2014} Izotov, Y. I, Thuan, T. X., \& Guseva, N. G. 2014, \textit{MNRAS}, 445, 778

\bibitem [Jacoby \& Ciardullo (1999)]{jacobyciardullo1999} Jacoby, G. H., \& Ciardullo, R. 1999, \textit{ApJ}, 515, 169

\bibitem [Jacoby \& Ford (1986)]{jacobyford1986} Jacoby, G. H., \& Ford, H. C. 1986, \textit{ApJ}, 304, 490

\bibitem [Karakas (2010)]{karakas2010} Karakas, A. I. 2010, \textit{MNRAS}, 403, 1413

\bibitem [Karakas \& Lattanzio (2003)]{karakaslattanzio2003} Karakas, A. I., \& Lattanzio, J. C. 2003, \textit{PASA}, 20, 393

\bibitem [Kniazev et al. (2014)]{kniazevetal2014} Kniazev, A. Y., Grebel, E. K., Zucker, D. B., et al. 2014, \textit{AJ}, 147, 16

\bibitem [Kwitter et al. (2012)]{kwitteretal2012} Kwitter, K. B., Lehman, E. M. M., Balick, B. et al. 2012, \textit{ApJ}, 753, 12

\bibitem [Magrini et al. (2009)]{magrinietal2009} Magrini, L., Stanghellini, L., \& Villaver, E. 2009, \textit{ApJ}, 696, 729

\bibitem [Peimbert et al. (2007)]{peimbertetal2007} Peimbert, M., Luridiana, V., \& Peimbert, A. 2007, \textit{ApJ}, 666, 636

\bibitem [Perinotto et al. (2004)]{perinottoetal2004} Perinotto, M., Morbidelli, L., \& Scatarzi, A. 2004, \textit{MNRAS}, 349, 793

\bibitem [Pignatari et al. (2013)]{pignatarietal2013} Pignatari, M, Herwig, F., Hirschi, R., et al. 2013, \textit{ApJS}, submitted

\bibitem [Richer \& McCall (2008)]{richermccall2008} Richer, M. G., \& McCall, M. L. 2008, \textit{ApJ}, 684, 1190

\bibitem [Richer \& McCall (2015)]{richermccall2015} Richer, M. G., \& McCall, M. L. 2015, unpublished

\bibitem [Roth et al. (2004)]{rothetal2004} Roth, M. M., Becker, T., Kelz, A. et al. 2004, \textit{ApJ}, 603, 531

\bibitem [Sanders et al. (2012)]{sandersetal2012} Sanders, N. E., Caldwell, N., McDowell, J., et al. 2012, \textit{ApJ}, 758, 133

\bibitem [Stanghellini et al. (2010)]{stanghellinietal2010} Stanghellini, L., Magrini, L., Villaver, E., \& Galli, D. 2010, \textit{A\&A}, 521, A3

\bibitem [Stanghellini et al. (2010)]{stanghellinietal2010} Stanghellini, L., Magrini, L., Casasola, V., \& Villaver, E. 2014, \textit{A\&A}, 567, A88

\bibitem [Stasi\'nska et al. (2013)]{stasinskaetal2013} Stasi\'nska, G., Pe\~na, M., Bresolin, F., \& Tsamis, Y. G. 2013, \textit{A\&A}, 552, A12

\end{thebibliography}
\end{document}